\documentclass[pre,twocolumn,twoside,showpacs,showkeys,floatfix]{revtex4}
\usepackage{dcolumn}%
\usepackage{bm}%
\usepackage{graphicx,textcomp}
\usepackage{amssymb,amsfonts,amsmath}
\usepackage{color}
\usepackage{nicefrac}

\newcolumntype{P}[1]{>{\centering\arraybackslash}p{#1}}
\newcolumntype{M}[1]{>{\centering\arraybackslash}m{#1}}

\begin{document}

\title{Coalescense with arbitrary--parameter kernels and monodisperse initial conditions: \\ A study within combinatorial framework}

\author{Micha\l{} \L{}epek\footnote{Corresponding author: lepek@if.pw.edu.pl}, Agata Fronczak, Piotr Fronczak}

\affiliation{Warsaw University of Technology, The Faculty of Physics, Koszykowa 75, PL-00-662 Warsaw, Poland}

\date{\today}



\begin{abstract}
For this work, we studied a finite system of discreet--size aggregating particles for two types of kernels with arbitrary parameters, a condensation (or branched--chain polymerization) kernel, $K(i,j)=(A+i)(A+j)$, and a linear combination of the constant and additive kernels, $K(i,j)=A+i+j$. They were solved under monodisperse initial conditions in the combinatorial approach where discreet time is counted as subsequent states of the system. A generating function method and Lagrange inversion were used for derivations. Expressions for an average number of clusters of a given size and its corresponding standard deviation were obtained and tested against numerical simulation. High precision of the theoretical predictions can be observed for a wide range of $A$ and coagulation stages, excepting post--gel phase in the case of the condensation kernel (a giant cluster presence is preserved). For appropriate $A$, these two kernels reproduced known results of the constant, additive and product kernels. Beside a previously solved linear--chain kernel, they extend the number of arbitrary--parameter kernels solved in the combinatorial approach.
\end{abstract}

\maketitle

\section{Introduction}\label{SectIntro}

Coagulation processes (also known as aggregation, agglomeration or coalescence) are amazingly common in nature. They determine a large number of phenomena, such as cloud formation, blood coagulation, milk curdling, traffic jam formation, and even planetary accretion. For decades, they were extensively investigated in several fields \cite{paper1, paper2, paper3, paper4, paper5, paper6, paper7,paper8,paper9,paper10}. A broad range of interdisciplinary applications can be found, including percolation phenomena in random graphs and complex networks \cite{2005_JPhysALushnikov, 2009_ScienceAchlioptas, 2010_PRLCosta, 2010_PRECho, 2016_PRLCho, 2016_Conv}, population genetics \cite{2005book_Hein}, pattern formation in social \cite{2014_PREMatsoukas, 2014_SciRepMatsoukas}, biological \cite{2007_EcolModelSaadi}, and man--made systems \cite{2002_PREDubovik}. Several technological applications are based on coagulation, including formation of polymers \cite{Stockmayer_1943}, food processing \cite{1981_Schmidt}, material processing \cite{Wattis_2004, Harris_2001}, and water treatment \cite{1993_Edzwald, paper11}. The aggregation formalism has recently started to play a significant role in modeling physiological processes \cite{2020_PRE_Nelson, 2020_Miangolarra}. Another interesting example of a broadly understood aggregation can be neurons accumulating into agglomerations observed in the neural network under specific conditions \cite{2018_ND_Lepek}.

In a coagulation process, particles merge irreversibly in a result of binary collisions (coagulation acts) which can be represented by the general scheme,
\begin{equation} \label{general_scheme}
   \left(i\right)+\left(j\right){{\stackrel{K\left(i,j\right)}{\longrightarrow}}}\left(i+j\right),
\end{equation}

\noindent where $\left(i\right)$ stands for a cluster of mass $i$ and $K\left(i,j\right)$ is the coagulation kernel representing the rate of the process. As the system is closed and clusters grow in time, eventually, all of the particles join to form a single giant cluster.

Deterministic (kinetic) methods for analysing coagulation processes base on the Smoluchowski aggregation equation \cite{paper12, paper13,paper14,paper15,paper16}. Requiring continuous cluster concentrations and infinite size of the system, the  Smoluchowski equation has been used to derive analytical solutions to few particular kernels, e.g., constant, $K(i,j)=const$, additive, $K(i,j) \propto i+j$, and multiplicative (product), $K(i,j) \propto ij$, and for selected initial conditions. This classic method uses a system of coupled nonlinear differential equations and provides mean--field evolution of the cluster size  distribution over time, not giving any information on the deviations from the mean solution. The equation provides an infinite--volume solution for the coagulation process and, as they are normalized with respect to the initial condition, they expire when the system moves away from the initial state. This results are failing for small systems and mature stages of the coagulation process. Therefore, explicit solutions to this equation are not ``exact solutions'' of the coagulation processes. Nevertheless, considerable literature exists on the existence and uniqueness of solutions to some general classes of kernels in Smoluchowski approach, including self--similar dynamical scaling solutions \cite{paper17,paper18,paper19,2004_Menon}. This area is still exhaustively researched \cite{Laurencot_2018, Bonacini_2019, Niethammer_2019, Banasiak_2019_Book, Canizo_2020}. On the basis of the Smoluchowski equation and the works of Ziff \cite{paper14}, a kinetic model of the gelling system consisting of two types of monomers was recently investigated \cite{2020_PRE_Nelson}. The Smoluchowski equation was also used to describe the process of protein aggregation in living cells \cite{2020_Miangolarra}.

The constant and the additive kernels are examples of the non--gelling kernels, in which the sol--gel transition is not observed when monodisperse initial conditions applied (in contrast to the product kernel). These three basic kernels are important as they became reference models of coagulation. The exact solutions to these kernels were obtained for the simplest case of monodisperse initial conditions as solutions to the master equation governing the time evolution of the probability distribution over system states \cite{paper20, 2004_PRLLushnikov, 2005_PRELushnikov, 2006_PhysD_Lushnikov, 2011_JPhysALushnikov} (stochastic approach), or through direct counting of these states \cite{grassberger1, grassberger2, 2018_PREFronczak, 2019_ROMP_Lepek} (combinatorial approach). An exact solution to the product kernel has been found for the case of arbitrary initial conditions \cite{2019_PRE_Fronczak}. The basic kernels were also revisited in a thermodynamic--focused description using the so--called linear ensembles \cite{2018_Matsoukas}.

Recently, the combinatorial approach proved its usefulness for solving more complex form of a kernel under monodisperse initial conditions, i.e., for the linear-chain kernel which included the electrorheological fluid coagulation process \cite{2021_PhysD_Lepek}. The solution involved arbitrary parameter $\alpha$ substantially increasing generality of the kernel form. The accuracy of theoretical solution tested versus numerical simulation and experimental data varied from approximate to ``exact'' depending on the parameter and coagulation time. It has been shown that for $\alpha=0$ the linear--chain kernel can be reduced to the constant kernel and reproduce its previously--known results.

\begin{center}
$\diamondsuit$    
\end{center}

The idea behind the combinatorial approach bases on the fact that successive steps of the coagulation process define the space of available states. The probability distribution over the state space can be determined by studying possible growth histories of clusters using combinatorial expressions. Then, the expressions for cluster size distribution and its standard deviation are derived. For that purpose, the model uses incomplete Bell polynomials and some involved combinatorics (please see \cite{2018_PREFronczak, 2019_ROMP_Lepek}). However, such an approach gave significant advantages. The first is an explicit expression for the number of clusters of a given size at a given aggregation step, valid for any kernel. The second advantage is an expression for the standard deviation of that average. It must be emphasized that these expressions does not change when adopted to cover any new type of kernel. The only problem that is needed to be addressed involves transforming a recurrent expression for the number of possible internal states of a cluster of a particular size to a non--recursive form.

The combinatorial framework requires several assumptions: (i) monodisperse initial conditions, (ii) discrete time (i.e., time steps counted by the subsequent states of the system), (iii) one coagulation act occurring at each time step, and (iv) relatively low particle concentrations since we do not consider the possibility of having encounters among 3 particles, or several encounters at a given time.

The condition (ii) needs a thorough comment. In a real coagulating system, particles merge (through binary collisions) more or less frequently. This frequency is determined by the kernel and by the evolution stage of the system. However, an exact time between these acts remains unknown (being stochastic in nature) and is modeled in, e.g., Marcus--Lushnikov framework of aggregation. In contrast, in the combinatorial approach we use, the time is counted in coagulation acts (steps). The first coagulation act occurs at time step $t=1$, the second coagulation act occurs at time step $t=2$, and so on. Therefore, that discrete time is no more than a counter of subsequent states of the system. It must be emphasized that this timescale of binary aggregations acts is not the same as the timescale of a physical time (measured, e.g., in seconds). Binary aggregation time steps are a (nonlinear) rescaled version of the physical time. Due to this reason, the combinatorial approach does not deal with the physical time and does not give information on ``how much time such a particular process would take in real''. The discrete combinatorial time must be regarded as time steps counting subsequent states of the system.

However, it is crucial that the lack of direct relevance to the physical time is not a serious limitation. If we have sufficient information on the real system: an initial number of clusters and a ``snapshot'' of the cluster size distribution in a particular point of time we can translate this real moment of the aggregation process to the language of discrete time steps. It is because the number of clusters in the system clearly defines the time step in the combinatorial formalism. We demonstrated such a procedure comparing combinatorial predictions to the experimental data in \cite{2021_PhysD_Lepek}.

Discrete time counted as subsequent aggregation acts has another important consequence. In general, the kernel, $K$, defines the relative probabilities of particular coagulation acts which are available to occur in the regarded system state (probability distribution of merging acts). In the combinatorial framework, strictly speaking, $K \propto f(i,j)$, so all kernel formulations can be modified by multiplying by any constant number which does not change the reaction rate of the process. For simplicity and consistency with mathematical tools, however, we will use the equal sign notation throughout the work.

\begin{center}
$\diamondsuit$    
\end{center}

For this work, using the combinatorial framework with recursive equations \cite{2019_ROMP_Lepek} we have studied two types of irreversible aggregation kernels with arbitrary parameters.

The first one is
\begin{equation} \label{condensation_kernel}
K\left(i,j\right) = (i+A)(j+A),
\end{equation}

\noindent where $A \ge 0$ stands for a parameter which can vary due to the choice of the particular process. According to D. Aldous \cite{paper10}, the kernel of this form represents condensation and branched--chain polymerization processes.

Condensation process is a particular case of coagulation which stands for the change of the physical state of matter from the gas phase into the liquid phase. Condensation has been researched in a number of fields, including (but not limited to) aerosol growth \cite{Drake_1972, Walter_1973, Pruppacher_1978, Wagner_1982}, water vapor condensation in cloud microphysics \cite{Long_1974, Gillespie_1975, Dziekan_2017}, and supersonic flows \cite{Wegener_1954, Yang_2017}.

In turn, branched--chain polymerization is a process of forming branched polymers which are defined as having secondary polymer chains linked to a primary backbone \cite{Stockmayer_1943}. There are several studies on the physical properties of polymers of this kind \cite{Gray_2009, Vilaplana_2010, Marshall_2013}. They are of special interest not only in industrial processing but also in biology \cite{Spouge_1983, Baskakov_2007, Fang_2011}.

The second kernel analysed in this work was the kernel combining constant and additive components in an arbitrary proportion,
\begin{equation} \label{combination_const_additve_kernel_with_B}
K\left(i,j\right) = A+B(i+j),
\end{equation}

\noindent with $A \ge 0$ and $B \ge 0$ being arbitrary parameters defining that proportion.

The kernel~(\ref{combination_const_additve_kernel_with_B}) is a part of the general kernel $ K = A + B (i + j) + Cij $, considered previosuly by Spouge \cite{Spouge_1983b}. Some stochastic solutions for the kernel $ K = A + B (i + j) $ were obtained and it was shown that for infinite systems, they are equivalent to kinetic solutions resulting from the Smoluchowski equation \cite{paper23}. In contrast to the present work, those solutions did not provide any information on the deviation from the mean behaviour, nor were validated against numerical results. 

In the combinatorial framework, as mentioned above, an exact formulation of the kernel~(\ref{combination_const_additve_kernel_with_B}) is $K \propto A+B(i+j)$. As such, its right--hand side can be easily normalized to obtain $B=1$, simplifying the formulation. Then, the resulting form of the kernel to be used for further calculations is
\begin{equation} \label{combination_const_additve_kernel}
K\left(i,j\right) = A+i+j.
\end{equation}

For $A=0$, Eq.~(\ref{combination_const_additve_kernel}) simply gives the additive kernel form. In turn, if $A \gg i,j$ then the reaction rate becomes relatively constant and ceases to depend on $i$ and $j$. Such a case correspond to the constant kernel. As the combinatorial approach assumes a finite volume system and monomeric initial conditions, the condition $A \gg i,j$ can also be regarded as a condition for $A$ to be much greater than the initial number of particles (i.e., than the number of monomer units in the system).

The paper is organized as follows. Section \ref{SecBull} presents the essentials of the combinatorial approach. Section \ref{SecDerivations} provides a detailed description of our method for calculating the number of possible internal states of a cluster for the condensation kernel and for the combination of the constant and additive kernels. It includes Lagrange inversion shown in detail. Section \ref{SecComp} compares theoretical predictions to the numerical simulations. Section \ref{SecSum} discusses relation to the system size, gives concluding remarks and describes possible extensions to this work.

\section{Bullet points of combinatorial formalism} \label{SecBull}

It may be beneficial for the Reader if we briefly describe essentials of the combinatorial approach to finite coagulating systems used in this work.

\subsection{Partial Bell polynomials}

A mathematical tool being a significant part of this formalism are partial Bell polynomials (also known as incomplete or the second kind of). They provide detailed information about the partition of an arbitrary set. They are defined as
\begin{multline} \label{Bell_polynom_def}
 B_{N,k}\left(a_1,a_2,\dots ,a_{N-k+1}\right)=B_{N,k}\left(\left\{a_s\right\}\right) \\ 
 = N!\sum_{\left\{n_s\right\}}{   \frac{a_1^{n_1} a_2^{n_2} \dots}{ n_1!n_2! \dots (1!)^{n_1} (2!)^{n_2} \dots }   } \\ = N!\sum_{\left\{n_s\right\}}{\prod^{N-k+1}_{s=1}{\frac{1}{n_s!}{\left(\frac{a_s}{s!}\right)}^{n_s}}},
\end{multline} 

\noindent where the summation is taken over all non--negative integers $\left\{n_s\right\}$ that satisfy
\begin{equation} \label{constraints}
 \sum^N_{s=1}{n_s=k} \;\;\;\; \textrm{and} \;\;\;\; \sum^N_{s=1}{sn_s=N}.
\end{equation}

Eq.~(\ref{Bell_polynom_def}) is a Diophantine equation, hence, in practice, other methods are needed to calculate values of the partial Bell polynomials (please see Appendix for an efficient recurrence equation).

\subsection{System state}

As mentioned in the Introduction, the combinatorial approach requires discrete time steps and monodisperse initial conditions (all of the clusters are monomers of size one). As a single coagulation act occurs at one time step, the total number of clusters, $k$, at time step $t$ is
\begin{equation} \label{EQ2}
 k=N-t,  
\end{equation}

\noindent where $N$ is the total number of monomeric units in the system. This number does not change during the coagulation process (preservation of mass) so the state of the system at time $t$ is described by Eq.~(\ref{constraints}). In this context, $n_s\ge 0$ stands for the number of clusters of mass $s$ ($s$ is the number of monomeric units which build the cluster) and $n_1$ corresponds to monomers, $n_2$ to dimers, $n_3$ to trimers, and so on. Therefore, a state of the system at time step $t$ is described by the sequence $\left\{ n_s \right\}$,
\begin{equation} \label{EQ3}
 \mathrm{\Omega }\left(t\right)=\left\{n_1,n_2,\dots ,n_s,\dots ,n_N\right\}.
\end{equation}

\subsection{Average number of clusters of given size and its standard deviation}

Being familiar with partial Bell polynomials, we can proceed to a brief outline of the main expressions arising from the combinatorial approach used here. They are applicable for any form of the kernel. They take their origins in three observations (for details and derivations, please refer to \cite{2018_PREFronczak, 2019_ROMP_Lepek}): (i) the set of monomers can be divided into subsets in a particular number of ways, (ii) coagulation acts for a particular cluster can be distributed in different time steps, and (iii) a given cluster could be created in a particular number of ways (i.e., the number of possible histories of a cluster) and this number must be derived for the kernel of interest. By combining these expressions together, one can calculate the average number of clusters, $\left\langle n_s\right\rangle$, of a given size, $s$, as
\begin{equation} \label{ns_general}  
 \left\langle n_s\right\rangle =\binom{N}{s}{\omega }_s\frac{B_{N-s,k-1}\left(\left\{{\omega }_g\right\}\right)}{B_{N,k}\left(\left\{{\omega }_g\right\}\right)},
\end{equation}

\noindent where, for simplicity,
\begin{equation} \label{omega_s}  
\omega_s = \frac{x_s}{(s-1)!} \;\;\;\; \textrm{and} \;\;\;\; \left\{\omega_g\right\} =  \left\{ \frac{x_g}{(g-1)!} \right\}.
\end{equation}

Eq.~(\ref{ns_general}) describes the average number of clusters of size $s$ after $t$ steps of the aggregation process (one coagulation acts occurs in one time step). Although $t$ is not explicitly present in the equation, $k$ plays $t$'s role, as $k$ is the total number of clusters in the system and decreases linearly with time.

In Eq.~(\ref{omega_s}), please note that $\omega_s$ is a single value and depends on cluster size $s$, while $\{\omega_g\}$ is a sequence not dependent on $s$, where $g$ varies from $1$ to $N-k+1$ (i.e., to $t+1$). In other words, $g$ is a maximal size of the cluster that \textit{could} arise in the system. The number $x_s$ (or $x_g$) used in Eq.~(\ref{omega_s}) can be interpreted as the number of possible internal states of a cluster of a given size. We will describe $x_s$ in detail further in this section.

As it was mentioned before, not only was the average number of clusters obtained in the combinatorial framework, but so was the corresponding standard deviation of this average,
\begin{equation} \label{std_dev_general}  
 {\sigma }_s=\sqrt{\left\langle n_s\left(n_s-1\right)\right\rangle +\left\langle n_s\right\rangle -{\left\langle n_s\right\rangle }^2}
\end{equation}

\noindent where
\begin{equation} \label{std_dev_general_addition}  
 \left\langle n_s\left(n_s-1\right)\right\rangle =\binom{N}{s,s}{{\omega }_s}^2\frac{B_{N-2s,k-2}\left(\left\{{\omega }_g\right\}\right)}{B_{N,k}\left(\left\{{\omega }_g\right\}\right)}\
\end{equation}

\noindent for $2s \leqslant N$, and $\left\langle n_s\left(n_s-1\right)\right\rangle = 0$ for other cases. For short, $\binom{N}{s,s}=\binom{N}{s}\binom{N-s}{s}$.

\subsection{Number of internal states of cluster of given size}

The last expression to be presented in this section is the expression for the number of internal states of a cluster of a given size, $x_s$ (or $x_g$). It can also be interpreted as the number of possible ways to create such a cluster (the number of its ''available histories''). Here, we will use subscript $g$ to emphasize that we study a single cluster of size $g$ that \textit{could} grow in the system. In previous works \cite{2019_ROMP_Lepek,2021_PhysD_Lepek}, it has been shown that this number can be calculated as
\begin{equation} \label{xg_recursive_definition_general}
 x_g = \frac{1}{2}\sum^{g-1}_{h=1}\binom{g}{h}\binom{g-2}{h-1}x_hx_{g-h}K(g,h),
\end{equation}

\noindent where $x_h$ and $x_{g-h}$ represent the numbers of ways to create two clusters of size $h$ and $(g-h)$ that joined and became a cluster of size $g$. $K(g,h)$ is an arbitrary kernel of our interest (translated into variables $g$ and $h$). The above recurrence expression is build as follows.

When the two clusters merge the resulting cluster of size $g$ appears. We can divide the cluster of size $g$ into subclusters of size $h$ and size $(g-h)$ in exactly $\binom{g}{h}$ ways. In other words, the first binomial factor denotes the number of ways of choosing a cluster of size $h$ out of $g$ monomers.

The second binomial factor, $\binom{g-2}{h-1}$, covers the fact that coagulation acts resulting in clusters $h$ and $(g-h)$ could occupy different possible time steps. We will use an example. There are two clusters: cluster of size $(g-h)=3$ and cluster of size $h=4$. To create these clusters we need five time steps. The cluster $(g-h)$ needs two time steps to arise, while the cluster $h$ needs three time steps to arise. Coagulation acts related to the creation of the cluster of size $h$ occupy $(h-1)$ time steps out of the total number of $(g-2)$ time steps needed to create clusters of sizes $h$ and $(g-h)$, hence, the binomial factor. Therefore, coagulation acts for the $(g-h)$ cluster and coagulation acts for the $h$ cluster could happen in ten different orders (sequences) in time, as $\binom{g-2}{h-1}=10$.

The sum in Eq.~(\ref{xg_recursive_definition_general}) is taken over all possible pairs of clusters that can join to become cluster of size $g$. The factor of $\frac{1}{2}$ is used to prevent double counting of the coagulation acts.

In the next sections, we will modify Eq.~(\ref{xg_recursive_definition_general}) to describe kernels to be solved and transform it to a non--recursive form that can be used in Eqs.~(\ref{ns_general})--(\ref{omega_s}) to calculate $\left\langle n_s\right\rangle$.

\section{Deriving solutions} \label{SecDerivations}

In this Section, we will derive the number $x_g$ for the kernels defined by Eqs.~(\ref{condensation_kernel}) and~(\ref{combination_const_additve_kernel}).

\subsection{Condensation kernel} \label{subsection_condensation}

At this point, we must translate the condensation kernel, $K\left(i,j\right) = \left( i+A \right) \left( j+A \right)$, into variables $g$ and $h$ used in Eq.~(\ref{xg_recursive_definition_general}). Parameters $i$ and $j$ are the masses of two merging clusters, $g$ is the mass of the resulting cluster, and $h$ is the mass of one of the subclusters (e.g., $i=h$). The condensation kernel can be rewritten as
\begin{equation} \label{K_gh_cond}
 K\left(g,h\right) = \left( h+A \right) \left( g-h+A \right).
\end{equation}

Then, the recursive expression for $x_g$ for the condensation kernel is
\begin{equation} \label{xg_recursive_condensation}
 x_g=\frac{1}{2}\sum^{g-1}_{h=1}\binom{g}{h}\binom{g-2}{h-1}x_hx_{g-h} \left( h+A \right) \left( g-h+A \right).
\end{equation}

The strategy of solving Eq.~(\ref{xg_recursive_condensation}) to obtain a non--recurrent expression for $x_g$ is based on the generating function method and is similar as in previous case of the simple kernels \cite{2019_ROMP_Lepek}.

Expanding binomial coefficients and substituting
\begin{equation} \label{condensation_subst}
y_g=\frac{x_g(g+A)}{g!(g-1)!}, 
\end{equation}

we obtain
\begin{equation} \label{condensation_2} 
 \frac{\left(g-1\right)}{(g+A)}y_g=\frac{1}{2}\sum^{g-1}_{h=1}{y_hy_{g-h}}.
\end{equation}

Now, we multiply both sides of Eq.~(\ref{condensation_2}) by $\sum^{\infty }_{g=1}{z^g}$ to have
\begin{equation} \label{condensation_3}  
 \sum^{\infty }_{g=1}{\frac{\left(g-1\right)}{(A+g)}y_gz^g}=\frac{1}{2}\sum^{\infty }_{g=1}{\sum^{g-1}_{h=1}{\left(y_hz^h\right)\left(y_{g-h}z^{g-h}\right)}}.
\end{equation}

The generating function is defined as $G\left(z\right)\equiv \sum^{\infty }_{g=1}{y_gz^g}$. Therefore, the right--hand side of Eq.~(\ref{condensation_3}) is equal to  $\frac{1}{2}G\left(z\right)G\left(z\right)$.

At this point, we must transform the left--hand side of Eq.~(\ref{condensation_3}) to the form containing $G\left(z\right)$. These transformations give
\begin{multline} \label{condensation_4}
 \sum^{\infty }_{g=1}{\frac{\left(g-1\right)}{(g+A)}y_gz^g} =
 \sum^{\infty }_{g=1}{{\frac{g}{g+A}}y_gz^g}-\sum^{\infty }_{g=1}{\frac{1}{g+A}y_gz^g} \\
 = \frac{1}{z^A}\sum^{\infty }_{g=1}{\frac{g}{g+A}y_gz^{g+A}}-\frac{1}{z^A}\sum^{\infty }_{g=1}{\frac{1}{g+A}y_gz^{g+A}} \\ = \frac{1}{z^A}\sum^{\infty }_{g=1}{gy_g \int z^{g+A-1} dz}-\frac{1}{z^A}\sum^{\infty }_{g=1}{y_g \int z^{g+A-1} dz} \\ 
 = \frac{1}{z^A}\int\sum^{\infty }_{g=1}{gy_g z^{g+A-1} dz}-\frac{1}{z^A}\int\sum^{\infty }_{g=1}{y_g z^{g+A-1} dz} \\ 
  = \frac{1}{z^A}\int z^A \sum^{\infty }_{g=1}{gy_g z^{g-1} dz}-\frac{1}{z^A} \int z^{A-1} \sum^{\infty }_{g=1}{y_g z^g}  dz \\ 
  = \frac{1}{z^A}\int z^A \sum^{\infty }_{g=1}{y_g \frac{\partial}{\partial z} \left( z^{g} \right) dz}-\frac{1}{z^A} \int z^{A-1} G(z) dz \\ 
  = \frac{1}{z^A}\int z^A  \frac{\partial}{\partial z} G(z) dz -\frac{1}{z^A} \int z^{A-1} G(z) dz.
 \end{multline}

As we transformed both sides of Eq.~(\ref{condensation_3}), it can be rewritten as
\begin{equation} \label{condensation_5}
\int z^A \frac{\partial }{\partial z} G(z) dz - \int z^{A-1} G(z) dz = \frac{1}{2}z^A G^2(z).
\end{equation}

In the next two equations, we will use $G$ instead of $G(z)$ for short. Differentiating both sides with respect to $z$ and after elementary transformations, we have
\begin{equation} \label{condensation_6}
\frac{ \partial G}{\partial z} = \frac{1}{z} \frac{\frac{A}{2}G^2+G}{1-G}
\end{equation}

\noindent which is an ordinary differential equation with separated variables. It can be solved for an arbitrary $A$ and its solution has the form of
\begin{equation} \label{condensation_7}
\frac{G}{\left(AG+2\right)^{\frac{A+2}{A}}} = Cz.
\end{equation}

Eq.~(\ref{condensation_7}) defines implicitly the generating function $G$. However, for further considerations, we need to have an explicit series representation of $G$. It can be derived using the Lagrange inversion method (see p.~148 in \cite{paper1}). Eq.~(\ref{condensation_7}) has a form of $f\left(G\right)=F$ where $f\left(G\right)$ stands for its left--hand side and $F$ stands for its right--hand side. Applying the Lagrange inversion to the equation of that form, the series representation of the inverse function $G\left(F\right)$ can be derived. Now, we will apply such a procedure to our problem.

In the Lagrange inversion, a general form of the series representation of the inverse function is given by
\begin{equation} \label{lagrange_inversion_general}
G(F) = a + \sum^{\infty}_{n=1}{G_n \frac{ \left( F-f(a) \right)^n }{ n! } },
\end{equation}

\noindent where 
\begin{equation} \label{lagrange_inversion_Gn}
G_n = \lim_{w \to a} \left[ \frac{d^{n-1}}{dw^{n-1}}  \left( \frac{w-a}{f(w)-f(a)} \right)^n  \right].
\end{equation}

Taking arbitrary $a=0$, we have $f(a=0)=0$, thus, considering Eq.~(\ref{condensation_7}), we can write
\begin{equation} \label{lagrange_inversion_a0}
G(F) = \sum^{\infty}_{n=1}{G_n \frac{F^n }{ n! } }
\end{equation}

\noindent and 
\begin{equation} \label{lagrange_inversion_Gn_ours}
G_n = \lim_{w \to 0} \left[ \frac{d^{n-1}}{dw^{n-1}}  \left( Aw+2 \right)^{\frac{2+A}{A}n}  \right].
\end{equation}

The technique for finding the sequence $G_n$ is to calculate values of Eq.~(\ref{lagrange_inversion_Gn_ours}) for the first few numbers $n$, e.g., $n=1,2,3,4$. Then, basing on these results, one shall guess the expression for $G_n$. Following this scheme, in our case, we can describe the sequence as
\begin{equation} \label{Gn_ours_guessed}
G_n = 2^{\frac{2n+A}{A}} [ 2n+nA ]^{*},
\end{equation}

\noindent where asterisk denotes a special product,
\begin{equation} \label{asterisk_function_cond}
[2n+nA]^{*}=\left\{
\begin{array}{ccc}
1 & \mbox{for} & n=1, \\
\prod^{n}_{m=2}{(2n+mA)} & \mbox{for} & n \ge 2.
\end{array}
\right.
\end{equation}

Finally, from the Lagrange inversion applied to Eq.~(\ref{condensation_7}), we obtained
\begin{equation} \label{lagrange_inversion_final}
G(z) = \sum^{\infty}_{g=1}{G_g \frac{ \left( Cz \right)^g }{ g! } },
\end{equation}

\noindent where $n$ was replaced by $g$ to stay consistent with our notation and $G_g$ is given by Eq.~(\ref{Gn_ours_guessed}). 

Until now, the constant $C$ remained unknown. At this point, we will calculate its value. Bearing in mind the definition of the generating function, $G(z) = \sum^{\infty }_{g=1}{y_gz^g}$, and considering Eq.~(\ref{lagrange_inversion_final}), we can write
\begin{equation} \label{comparison_to_definition}
\sum^{\infty}_{g=1}{G_g \frac{ \left( Cz \right)^g }{ g! } } = \sum^{\infty }_{g=1}{y_gz^g}.
\end{equation}

Comparing the first elements of these sums we have $G_1 \frac{ Cz }{ 1!} = {y_1 z}$. Therefore,
\begin{equation} \label{comparison_first_coefficients}
G_1 C = y_1.
\end{equation}

From the initial condition, $x_1 \equiv 1$, and considering the substitution, Eq.~(\ref{condensation_subst}), we obtain $y_1=A+1$. On the other hand, $G_1$ is given by Eq.~(\ref{Gn_ours_guessed}). In consequence,
\begin{equation} \label{constant_C}
C = 2^{-\frac{A+2}{A}}(A+1).
\end{equation}

Now, considering the respective coefficients in Eq.~(\ref{comparison_to_definition}), $G_g \frac{ C^g }{ g! } = {y_g}$, and after elementary transformations, we have
\begin{equation} \label{yg_final_cond}
y_g = \frac{(A+1)^g}{2^{g-1}g!} [2g+gA]^{*} .
\end{equation}

The last step to derive $x_g$ is going back from the substitution, Eq.~(\ref{condensation_subst}). Finally, we obtain the number of ways to create a cluster of size $g$ for the condensation kernel as
\begin{equation} \label{xg_final_cond}
x_g = \frac{(A+1)^g (g-1)!}{2^{g-1}(A+g)} [2g+gA]^{*},
\end{equation}

\noindent where the special function, $[ . ]^*$, is given by Eq.~(\ref{asterisk_function_cond}).

Eq.~(\ref{xg_final_cond}) together with Eqs.~(\ref{ns_general}) and~(\ref{omega_s}) fully defines the average number of clusters of a given size at a given time step, $\langle n_s \rangle$, for the condensation kernel.

\subsection{Combination of constant and additive kernels}

In this Section, we will derive the number of possible internal states of a cluster of given size $g$, $x_g$, for the kernel (\ref{combination_const_additve_kernel}). 

Similarly as in the case of the condensation kernel, we must translate the kernel of interest, this time $K\left(i,j\right) =  A+i+j$, to the language of $g$ and $h$ used in the general expression for $x_g$, Eq.~(\ref{xg_recursive_definition_general}). Bearing in mind that $i$ and $j$ are the masses of two merging clusters, $g$ is the mass of the resulting cluster and $h$ is the mass of one of the sub-clusters (e.g., $i=h$), we can rewrite the kernel as
\begin{equation} \label{K_gh_comb}
 K\left(g,h\right) = A + h + (g-h) = A + g.
\end{equation}

Hence, using Eq.~(\ref{xg_recursive_definition_general}) the recurrent expression for $x_g$ for this kernel is
\begin{equation} \label{xg_recurrent}
 x_g=\frac{1}{2}\sum^{g-1}_{h=1}\binom{g}{h}\binom{g-2}{h-1}x_hx_{g-h} \left( A + g \right).
\end{equation}

Expanding binomial factors and substituting
\begin{equation} \label{substitution} y_g = \frac{x_g}{g!(g-1)!}, 
\end{equation}

\noindent we obtain
\begin{equation} \label{combination_2} 
 \frac{\left(g-1\right)}{(g+A)}y_g=\frac{1}{2}\sum^{g-1}_{h=1}{y_hy_{g-h}}.
\end{equation}

Eq.~(\ref{combination_2}) has the same form as the analogous equation for the condensation kernel with the only difference in the substitution, Eq.~(\ref{substitution}). Thus, we will use the solution from \ref{subsection_condensation}. However, because of the substitution we need to calculate $C$ once again. Now, from the initial condition ($x_1 \equiv 1$), we have $y_1=1$, so the constant is
\begin{equation} \label{constant_C_again}
C = 2^{-\frac{A+2}{A}}.
\end{equation}

Therefore,
\begin{equation} \label{yg_final_comb}
y_g = \frac{1}{2^{g-1}g!} [2g+gA]^{*},
\end{equation}

\noindent where the asterisk denotes a special product given by Eq.~(\ref{asterisk_function_cond}).

Finally, going back from the substitution, Eq.~(\ref{substitution}), we obtain the number of ways to create a cluster of size $g$ for the kernel (\ref{combination_const_additve_kernel}) as
\begin{equation} \label{xg_final_comb}
x_g = \frac{(g-1)!}{2^{g-1}} [2g+gA]^{*},
\end{equation}

\noindent where, again, the product function, $[ . ]^*$, is given by Eq.~(\ref{asterisk_function_cond}).

Eq.~(\ref{xg_final_comb}) together with Eqs.~(\ref{ns_general}) and~(\ref{omega_s}) fully defines the average number of clusters of a given size at a given time step, $\langle n_s \rangle$, for the combination of the constant and additive kernels.

\section{Theoretical results compared to numerical simulations} \label{SecComp}

\begin{figure*}[ht!] 
\includegraphics[width=1\textwidth]{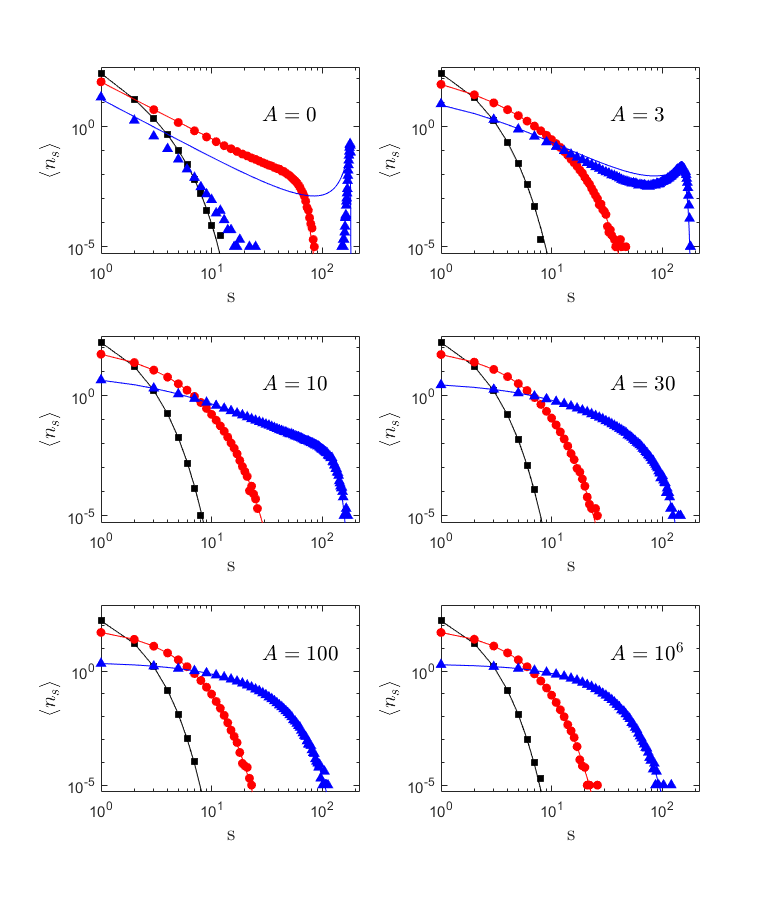}
\caption{Theoretical calculations of the average number of clusters of size $s$, $\langle n_s \rangle$, for the condensation kernel, $K(i,j)=(A+i)(A+j)$, with different values of $A=\{0, 3, 10, 30, 100, 10^6 \}$ compared to the numerical simulation results. Solid lines represent theoretical predictions based on the combinatorial equations, Eqs.~(\ref{ns_general_final})--(\ref{omega_g_condensation_final}).  Although theoretical results are defined only for integer $s$, solid lines are used as guidelines for eyes. Circles, squares and triangles correspond to the results obtained by numerical simulation. The number of monomers in the system was $N=200$. Three stages of the coagulation process are presented: $t=20$ (beginning of the process, $\nicefrac{t}{N}=0.1$, squares, black), $t=100$ (the half-time, $\nicefrac{t}{N}=0.5$, circles, red) and $t=180$ (late stage of the process $\nicefrac{t}{N}=0.9$, triangles, blue). For $A=0$, the results (both theoretical and numerical) fully correspond to the results obtained for the product kernel. For $A \gg N$ (here, $A=10^6$), the results correspond accurately to the known results for the constant kernel. For each case, ${10}^5$ independent simulations were performed.}
\label{Figure_condensation}
\end{figure*}

\begin{figure*}[ht!] 
\includegraphics[width=1\textwidth]{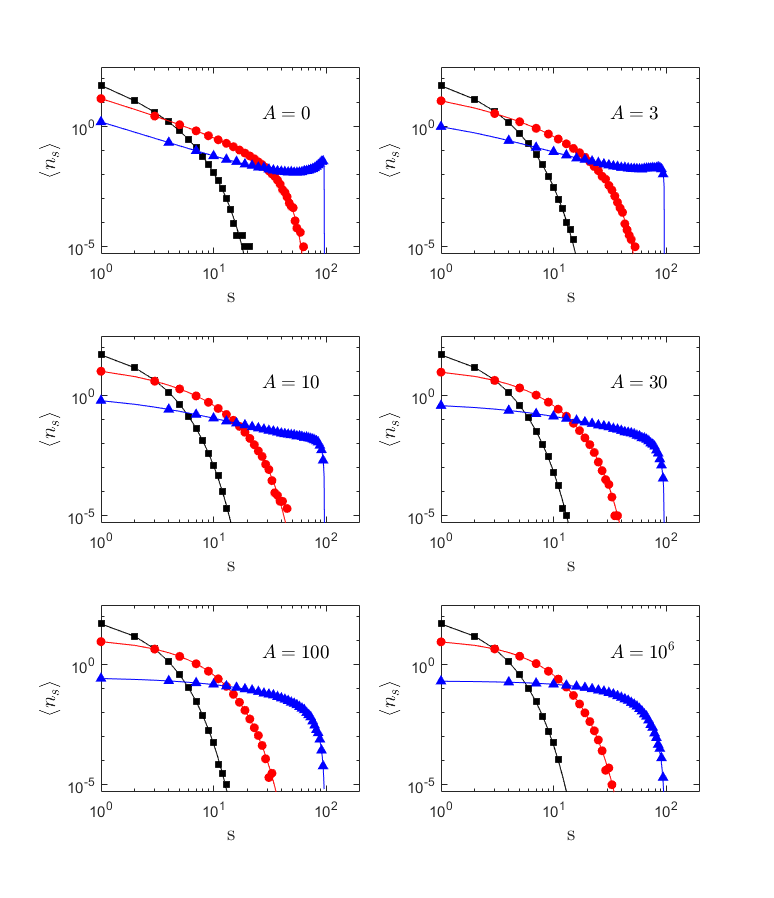}
\caption{Theoretical calculations of the average number of clusters of a given size, $\langle n_s \rangle$, for the combination of constant and additive kernels, $K(i,j)=A+i+j$, with several values of $A=\{0,3,10,30,100,10^6\}$, compared to the numerical simulation results. Solid lines represent theoretical predictions based on the combinatorial expressions, Eq.~(\ref{ns_general_final_comb})--(\ref{omega_g_combination_final}). Although theoretical results are defined only for integer $s$, solid lines are used as guidelines for eyes. Circles, squares and triangles correspond to the results obtained by the numerical simulation. The number of monomers in the system was $N=100$. Three stages of the coagulation process are presented: $t=30$ (beginning stage of the process, $\nicefrac{t}{N}=0.3$, squares, black), $t=70$ (moderate time, $\nicefrac{t}{N}=0.7$, circles, red) and $t=95$ (a very late stage of the process, $\nicefrac{t}{N}=0.95$, triangles, blue). Subsequent values of $A$ represent different proportions of the constant and additive parts involved in the kernel. For $A=0$, the results (both theoretical and numerical) fully correspond to the results obtained previously for the additive kernel. For $A \gg N$ (here, $A=10^6$), the results correspond accurately to the known results for the constant kernel. For each case, ${10}^5$ independent simulations were performed.}
\label{Figure_combination}
\end{figure*}

\begin{figure*}[ht!] 
\includegraphics[width=1\textwidth]{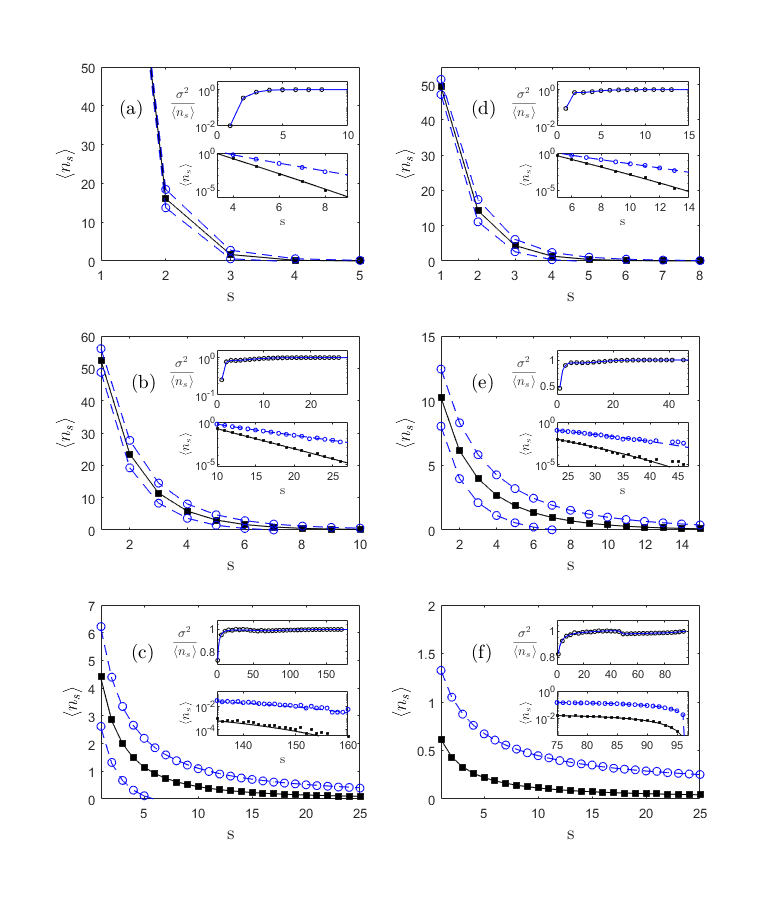}
\caption{Standard deviation predicted by the theoretical model, Eqs.~(\ref{std_dev_general}) and~(\ref{std_dev_general_addition}), vs. standard deviation obtained by the simulation for the condensation kernel and for the combination of the constant and additive kernels. For both, $A=10$. Solid and dashed lines represent combinatorial results for the average number of clusters of a given size, $\langle n_s\rangle$, and for its standard deviation, respectively; squares and circles represent, respectively, the average number of clusters and standard deviation obtained by simulation. Left column: the condensation kernel, $N=200$, (a) $t=20$, $\nicefrac{t}{N}=0.1$, (b) $t=100$, $\nicefrac{t}{N}=0.5$, and (c) $t=180$, $\nicefrac{t}{N}=0.9$. Right column: the combination of the constant and additive kernels (for this case, $N=100$), (d) $t=30$, $\nicefrac{t}{N}=0.3$, (e) $t=70$, $\nicefrac{t}{N}=0.7$, and (f) $t=95$, $\nicefrac{t}{N}=0.95$. Upper inset figures: plots of variance divided by mean, $ \sigma^2/ \langle n_g\rangle$. Again, solid lines for theoretical predictions and circles for numerical data. Lower inset figures: plots of $\langle n_s\rangle$ for the highest $s$ (logarithmic scale). For each case, ${10}^5$ independent simulations were performed.}
\label{Figure_stddev}
\end{figure*}

In this Section, we will compare theoretical solutions arising from the combinatorial framework which were derived in Section~\ref{SecDerivations} to numerical results obtained by the simulation.

The numerical simulation term refers to a simulation of the aggregating system of $N$ particles where the information on the number of clusters and on the masses of these clusters are stored in the program memory and updated step by step in order with subsequent aggregation events. For each aggregation event, two clusters are selected randomly from the distribution suitable for the kernel which is simulated. The crucial part of the simulation is implementing the reaction kernel, $K$. In the previous work on the linear--chain kernel, an exhaustive description on how to implement an arbitrary kernel in such a simulation has been given (Appendix in \cite{2021_PhysD_Lepek}). The code used for the simulations, as well as for theoretical calculations, is available in the Internet \cite{cpp_libraries}.

\subsection{Final expressions for condensation kernel}

To have a final expression for the average number of clusters of a given size at a given time step, $\left< n_s \right>$, we need to use Eqs.~(\ref{ns_general}) and~(\ref{omega_s}) with $x_g$ specified by Eqs.~(\ref{xg_final_cond}) and~(\ref{asterisk_function_cond}). In effect, we obtain a set of self--contained expressions to calculate $\left< n_s \right>$ for the condensation kernel, $K(i,j)=(A+i)(A+j)$. For clarity, we will write these expressions explicitly, 
\begin{equation} \label{ns_general_final}  
 \left\langle n_s\right\rangle =\binom{N}{s}{\omega }_s\frac{B_{N-s,k-1}\left(\left\{{\omega }_g\right\}\right)}{B_{N,k}\left(\left\{{\omega }_g\right\}\right)},
\end{equation}

\noindent where the time is counted by the total number of clusters remaining in the system, $k=N-t$, and
\begin{equation} \label{omega_s_condensation_final}  
\omega_s = \frac{(A+1)^s}{2^{s-1}(A+s)} [2s+sA]^{*},
\end{equation}
\begin{equation} \label{omega_g_condensation_final}  
\left\{\omega_g\right\} =  \left\{ \frac{(A+1)^g}{2^{g-1}(A+g)} [2g+gA]^{*} \right\}
\end{equation}

\noindent and $[ . ]^*$ denotes the special product function given by
\begin{equation} \label{asterisk_function_cond_final}
[2g+gA]^{*}=\left\{
\begin{array}{ccc}
1 & \mbox{for} & g=1, \\
\prod^{g}_{h=2}{(2g+hA)} & \mbox{for} & g \ge 2.
\end{array}
\right.
\end{equation}

Using the above set of equations we can calculate cluster size distribution for any stage of the system's evolution. The only parameter here is an arbitrary real value $A \ge 0$.

Taking advantage of a known relation for Bell polynomials \cite{Comtet_1974},
\begin{multline} \label{Bell_relation_1}  
B_{N,k}\left(abz_1,ab^2z_2,\dots,ab^{N-k+1}z_{N-k+1}\right) \\ = 
 a^kb^NB_{N,k}\left(z_1,z_2,\dots,z_{N-k+1}\right),
\end{multline}

\noindent we can simplify the set of Eqs.~(\ref{ns_general_final})--(\ref{omega_g_condensation_final}) to the form of
\begin{equation} \label{ns_general_final_simplified}  
 \left\langle n_s\right\rangle =\binom{N}{s} \frac{[2s+sA]^{*}}{(A+s)} \frac{B_{N-s,k-1}\left(\left\{ \frac{[2g+gA]^{*}}{A+g} \right\}\right)}{B_{N,k}\left(\left\{\frac{[2g+gA]^{*}}{A+g} \right\}\right)}.
\end{equation}

In Figure~\ref{Figure_condensation}, the theoretical solutions arising from our combinatorial equations were plotted against the results of the numerical simulation for several values of $A=\{0,3,10,30,100,10^6\}$ to show the performance of the solutions in respect to $A$. The total number of monomers in the system was $N=200$. For each $A$, we present three stages of the aggregation process. These are: $t=20$ (when the aggregation begins, $\nicefrac{t}{N}=0.1$), $t=100$ (half--time, $\nicefrac{t}{N}=0.5$), and $t=180$ (a late stage of the process, $\nicefrac{t}{N}=0.9$). The case of $A=0$ corresponds to the product kernel which is the widely--known example of a gelling kernel. It can be observed that, for $t=20$ and $t=100$, the theoretical predictions model the results of the simulation with a high precision. For the latest stage of the process, this accuracy significantly falls with the theoretical curve only preserving the giant cluster appearing in the system. The results for $A=0$, both theoretical and numerical, fully correspond to the results obtained previously for the product kernel in \cite{2019_ROMP_Lepek}. In the case of $A=3$ and the latest time, $t=180$, the theoretical solution behaves similarly to the numerical data, thus, the combinatorial solution can be regarded as approximate. For $A=3$ and earlier times, the theoretical results are perfectly accurate. As well, for higher values of $A$ the coagulating system is modeled with an excellent precision for any time of the process. For $A \gg i,j$, the solution (so the numerical simulation) converges to the case of the constant kernel.

Additionally, using Eqs.~(\ref{std_dev_general}) and~(\ref{std_dev_general_addition}) we can easily obtain standard deviation of the average numbers of clusters. For the condensation kernel, these theoretical predictions of the standard deviation were compared to the numerically--obtained results in Figure~\ref{Figure_stddev}.

\subsection{Final expressions for combination of constant and additive kernels}

Again, to obtain a final expression for the average number of clusters of a given size, $\left< n_s \right>$, we need to use Eqs.~(\ref{ns_general}) and~(\ref{omega_s}) with $x_g$ specified by Eqs.~(\ref{xg_final_comb}) and~(\ref{asterisk_function_cond}). Eventually, we have a set of self--contained expressions for the kernel of the form $K(i,j)=A+i+j$. Once again, let us write these expressions explicitly,
\begin{equation} \label{ns_general_final_comb}  
 \left\langle n_s\right\rangle =\binom{N}{s}{\omega }_s\frac{B_{N-s,k-1}\left(\left\{{\omega }_g\right\}\right)}{B_{N,k}\left(\left\{{\omega }_g\right\}\right)},
\end{equation}

\noindent where
\begin{equation} \label{omega_s_combination_final}  
\omega_s = \frac{1}{2^{s-1}} [2s+sA]^{*},
\end{equation}
\begin{equation} \label{omega_g_combination_final}  
\left\{\omega_g\right\} =  \left\{ \frac{1}{2^{g-1}} [2g+gA]^{*} \right\},
\end{equation}

\noindent and $[ . ]^*$ denotes, again, the special function given by Eq.~(\ref{asterisk_function_cond_final}). Analogously as for the condensation kernel, we can calculate cluster size distribution for an arbitrary real parameter $A \ge 0$ at arbitrary time of the system evolution.

Again, using the relation (\ref{Bell_relation_1}), the set of Eqs.~(\ref{ns_general_final_comb})--(\ref{omega_g_combination_final}) can be simplified to the form of
\begin{equation} \label{ns_general_final_comb_simplified}  
 \left\langle n_s\right\rangle =\binom{N}{s} [2s+sA]^{*} \frac{B_{N-s,k-1}\left(\left\{ [2g+gA]^{*} \right\}\right)} {B_{N,k}\left(\left\{[2g+gA]^{*} \right\}\right)}.
\end{equation}

In Figure~\ref{Figure_combination}, average numbers of clusters of a given size, $\left< n_s \right>$, are presented for several values of $A=\{0,3,10,30,100,10^6\}$. There are three data series contained in each plot, corresponding to three different stages of the system evolution. The initial number of monomers, thence, the total number of monomeric units in the system was $N=100$. The first series ($t=30$, $\nicefrac{t}{N}=0.3$) corresponds to the early stage of the coagulation process. The second series ($t=70$, $\nicefrac{t}{N}=0.7$) corresponds to the later stage of the process. The third series presents the cluster size distribution at the end of the process ($t=95$, $\nicefrac{t}{N}=0.95$), only five steps before the moment when all of the particles join into one single cluster. For $A=0$, the results (both theoretical and numerical) fully correspond to the results obtained previously for the additive kernel \cite{2019_ROMP_Lepek}. For $A \gg N$, the process can be regarded as usual constant kernel coagulation. In Figure~\ref{Figure_combination}, we have used $A=10^6$ and the results correspond accurately to the known result for the constant kernel. For all values of $A$, as well as for any time, the combinatorial predictions proved to be ``exact'' solutions of the process.

In Figure~\ref{Figure_stddev}, we present standard deviation of the cluster size distribution given by the combinatorial approach, Eqs.~(\ref{std_dev_general}) and~(\ref{std_dev_general_addition}), for $A=10$ and several stages of the system evolution, $t=30,70,95$. We compared them to the standard deviation calculated for the data obtained by the simulation. Combinatorial estimates followed numerical deviation with perfect precision for any time of the process.

\begin{figure}[ht]
\includegraphics[scale=0.6]{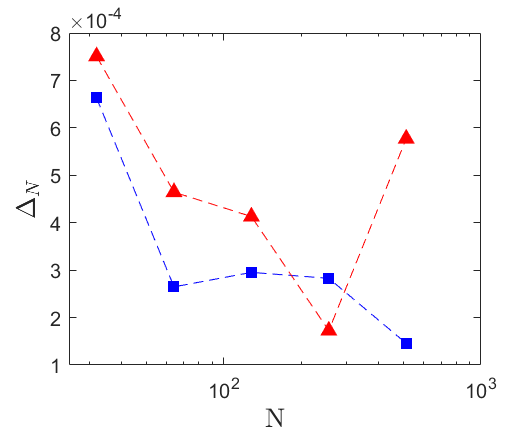}
\caption{Absolute difference between theoretical predictions and numerical data normalized by system size, $\Delta_N = \frac{1}{N} \sum_s{ | \langle n_s \rangle^{theory} - \langle n_s \rangle^{simul} | }$, for $N = \{  2^5, 2^6, 2^7, 2^8, 2^9 \}$. Blue squares represent condensation kernel, red triangles represent combination of the constant and additive kernels. Dashed lines are only guidelines for eyes. For both kernels, $A=10$ and $\nicefrac{t}{N}=0.5$ were used. For each case, $10^5$ independent simulations were performed, excepting $N=2^9$ when it was $10^4$.}
\label{Figure_sizedep}
\end{figure}

\begin{figure*}[ht]
\includegraphics[width=0.85\textwidth]{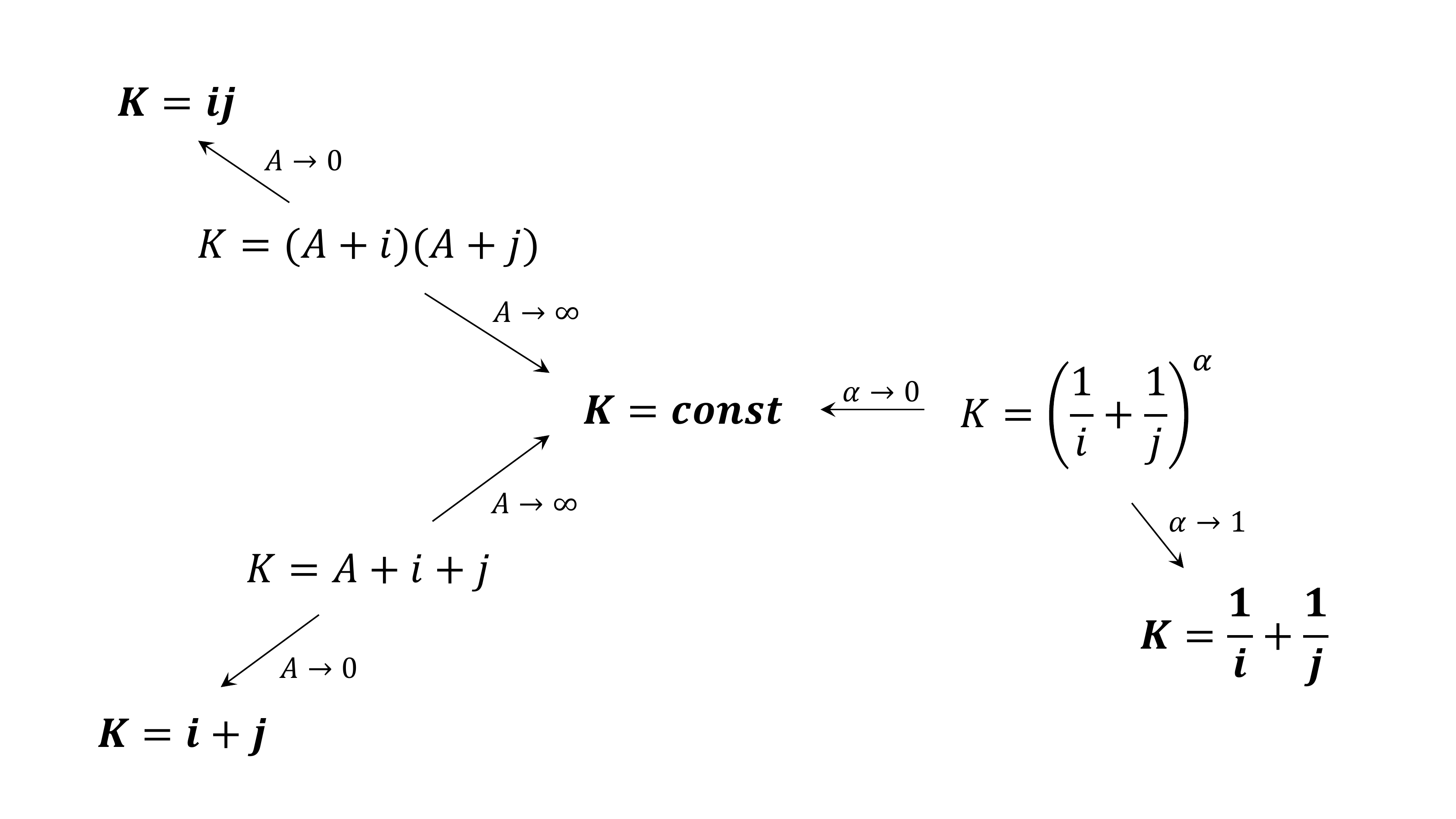}
\caption{Some of the kernel forms solved in the combinatorial approach can be reduced to a form corresponding to the constant kernel. These are following kernels: previously solved linear--chain kernel, $ K = \left(i^{-1} + j^{-1} \right)^{\alpha} $, condensation kernel, $ K=(A+i)(A+j) $, and a linear combination of the constant and additive kernels, $ K = A + i + j $. In the case of the latter two, for the constant value $ A =0 $, one obtains previously known kernels, respectively, product and additive. In this way, theoretical solutions were obtained for all "intermediate" processes between the constant kernel and the product, additive, and linear--chain kernels.}
\label{Figure_Summary}
\end{figure*}

\section{Discussion and conclusion} \label{SecSum}



For this work, using combinatorial framework with recursive equations we have studied the two types of irreversible aggregation kernels with arbitrary parameters, the condensation kernel, $K(i,j) = (A+i)(A+j)$, and the combination of the constant and additive kernels, $K(i,j)=A+i+j$. Expressions for the average number of clusters of a given size and for the corresponding standard deviation were obtained and plotted against the numerical simulation results.

What are the advantages of the above analysis over a conventional solutions of the rate equations? As far as we know there were no “exact” solution of the rate equation with the condensation kernel. On the other hand, some stochastic solutions for the kernel $ K = A + B (i + j) $ were obtained \cite{Spouge_1983b} and it was shown that, for infinite systems, they are equivalent to kinetic solutions of the Smoluchowski equation \cite{paper23}. In contrast to the present work, they did not provide any information on the deviation from the mean behaviour, nor were validated by the numerical results. Additionally, it is worth to note that the discreteness of the combinatorial approach does not diminish the
generality of the solutions as it allows to obtain continuous--time results by the asymptotic analysis (see, e.g., Eqs.~(60)--(62) in \cite{2018_PREFronczak}).

While presenting results in the Figures we used rather small sizes of the systems. However, the combinatorial framework allows to analyze systems of any size, including large ones. The problem is that obtaining relevant numerical simulation results for the size larger than $N = 400$ particles for an arbitrary kernel is computationally expensive due to the fact that we need to repeat the simulation millions of times to obtain reliable statistics. This is the main reason why we did not compare combinatorial and numerical results for larger sizes. To investigate possible dependency on the size and to be sure that quality of the theoretical predictions does not decrease with increasing system size we performed additional simulations for several values of $N$. In Figure~(\ref{Figure_sizedep}), we present an absolute difference between numerical and theoretical results as a function of the system size. No alarming observations were made.

Computational complexity of the expressions including Bell polynomials also may be an issue for calculations. However, in some cases (e.g., electrorheological coagulation \cite{2021_PhysD_Lepek}), it was possible to reduce Bell polynomials to much simpler form. Then, one can easily analyze much larger system sizes. Additionally, an efficient formula to calculate partial Bell polynomials is provided in the Appendix. Nonetheless, we suppose computational issues to be diminishing in future due to the increasing computational power of machines.

Some of the kernel forms solved in the combinatorial approach so far, i.e., linear--chain, $ K = \left(i^{-1} + j^{-1} \right)^{\alpha} $, condensation, $ K = (A + i) ( A + j) $, and a linear combination of the constant and additive kernels, $ K = A + i + j $, can be reduced to a form corresponding to the constant kernel. For the latter two and $ A=0 $, we obtain the previously known kernels, respectively, product and additive. In this way, theoretical solutions were obtained for all "intermediate" processes between the constant kernel and the product, additive, and linear--chain kernels (as shown in the Figure~\ref{Figure_Summary}). However, the quality of the combinatorial solutions decreases with increasing $ \alpha $. In the case of a condensation kernel, accuracy of the combinatorial solutions decreases once the gel point is exceeded. For the combination of the constant and additive kernels, the combinatorial solutions remain ''exact'' for any $ A $ and for any stage of the aggregation. Obtaining explicit theoretical solutions for such a wide continuous space of kernel forms can be perceived as a novelty in a research on the coagulation processes.


Obviously, several questions still remain unanswered. Why do combinatorial expressions give exact results for some kernels while approximate for other? Is it possible to modify the equations to cover arbitrary initial conditions or the post--gel phase as it was done in the for the product kernel in the Marcus--Lushnikov approach \cite{2019_PRE_Fronczak}? Are there other kernels which can be solved in the combinatorial framework with arbitrary parameters and extend the view in Figure~\ref{Figure_Summary}? Answering any of these questions would be a serious step forward in the coagulation theory.

\acknowledgments
This work has been supported by the National Science Centre of Poland (Narodowe Centrum Nauki) under grant no.~2015/18/E/ST2/00560 (A.F. and M.\L.).

\section*{Appendix}

As incomplete Bell polynomial definition is given by a Diophantine equation (which is NP--hard problem) other expression must be used to calculate its value in practice. An efficient way to calculate these polynomials is recurrence relation,
\begin{multline} \label{Bell_recursion}  
 B_{n,k} \left( a_1, a_2, \dots, a_m, \dots, a_{n-k+1} \right) \\ = \sum_{m=1}^{n-k+1} \binom{n-1}{m-1} a_{m} B_{n-m,k-1},
\end{multline}
 
\noindent where $B_{0,0} = 1$, $B_{n,0} = 0$ for $n\ge1$, and $B_{0,k} = 0$ for $k\ge1$.

Some computational environments implement Bell polynomials (e.g., Wolfram Mathematica). For some others, a relevant code can be easily found in the Internet (e.g., Matlab). A problem in applying theoretical predictions is the explosion of digits of Bell polynomials for $N>70$ as this number immediately exceeds the precision available in standard programming environments. Therefore, calculations need to be performed with the help of arbitrary precision computation packages. An implementation for C++ environment using an arbitrary precision package can be found in \cite{cpp_libraries}.

\end{document}